\documentclass[prc,twocolumn,superscriptaddress,showpacs,preprintnumbers,amsmath,amssymb,floatfix]{revtex4}

\usepackage{graphicx}
\usepackage{dcolumn}
\usepackage{bm}

\begin{document}


\title{Effective interaction for $pf$-shell nuclei}

\author{M. Honma}
\affiliation{
Center for Mathematical Sciences, University of Aizu,
 Tsuruga, Ikki-machi, Aizu-Wakamatsu, Fukushima 965-8580, Japan}
\author{T. Otsuka}
\affiliation{
Department of Physics, University of Tokyo, Hongo, Tokyo 113-0033, Japan}
\affiliation{
RIKEN, Hirosawa, Wako-shi, Saitama 351-0198, Japan}
\author{B. A. Brown}
\affiliation{
National Superconducting Cyclotron Laboratory and Department 
of Physics and Astronomy,
Michigan State University, East Lansing, MI 48824-1321}
\author{T. Mizusaki}
\affiliation{
Institute of Natural Sciences, Senshu University, Higashimita, 
Tama, Kawasaki, Kanagawa 214-8580, Japan}

\date{\today}

\begin{abstract}
An effective interaction is derived for use in the full $pf$ basis.
Starting from a realistic G-matrix interaction,
 195 two-body matrix elements and 4 single-particle energies are determined 
 by fitting to
 699 energy data in the mass range 47 to 66.
The derived interaction successfully describes
various structures of $pf$-shell nuclei.
As examples, systematics of the energies of the first $2^+$ states in
the Ca, Ti, Cr, Fe, and Ni isotope chains and energy levels of $^{56,57,58}$Ni
 are presented. The appearance of a new magic number 34 is seen.
\end{abstract}


\maketitle

The nuclear shell model has been very successful in our understanding
 of nuclear structure:
 once a suitable effective interaction is found,
 the shell model can reproduce/predict various observables
 accurately and systematically.
For light nuclei, there are several ``standard'' effective interactions
 such as the Cohen-Kurath \cite{ck} and the USD \cite{usd} interactions
 for the $p$ and $sd$ shells, respectively.
On the other hand, in the next major shell, i.e., in the $pf$-shell,
a unified effective interaction for all nuclei in this region
has not been available.
The $pf$-shell is quite important for a variety of problems
in nuclear structure, such as the 
stability/softness of the magic number 28, and nuclear 
astrophysics, such
as electron capture in supernovae explosions.  Thus, a sound,
systematic and precise description of the $pf$-shell nuclei is
urgent and important.
In this Letter, we present a unified effective interaction 
which one can 
apply to the shell-model description of the entire $pf$-shell.

The spin-orbit splitting gives rise to
 a sizable energy gap in the $pf$-shell between the $f_{7/2}$ orbit 
 and the other orbits ($p_{3/2}$, $p_{1/2}$, $f_{5/2}$),
producing the $N$ or $Z$=28 magic number.
However, the excitations across the gap are important for
ground and excited state properties of many $pf$-shell nuclei.
It is intriguing to understand how 
this magic number persists or 
fades away in various situations.
We shall discuss this point in this Letter, and the word ``cross-shell'' 
refers to the $N$ or $Z$=28 shell gap hereafter.  
Because of such cross-shell mixing, shell-model 
calculations including all $pf$-shell configurations are 
necessary, and the predictive power obtained with a unified 
effective interaction in the full $pf$-shell space is very
important.
The full $pf$-shell calculation, however, leads to 
the diagonalization of huge Hamiltonian matrices with dimensions
of up to 2 billion. The extreme difficulty of dealing with such
large matrices is the major
reason why a unified effective interaction has been missing for the
$pf$ shell.  
The Monte Carlo Shell Model (MCSM) introduced recently
\cite{qmcd,mcsm} has changed the situation by making such 
calculations feasible over the entire region of the $pf$-shell.
Also conventional shell-model calculations have advanced.

The effective interaction can in principle be derived from the free 
nucleon-nucleon interaction.
In fact such microscopic interactions have been proposed
for the $pf$-shell \cite{kb,g-mat} with certain success particularly 
in the beginning of the shell.
These interactions, however, fail
 in cases of many valence nucleons, e.g., 
$^{48}$Ca \cite{g-mat} and $^{56}$Ni.

These microscopic interactions
 can be modified empirically so as to better 
reproduce experimental data.
The monopole-modified interaction, KB3 \cite{kb3},
and also the shell-gap readjusted version, KB3G\cite{kb3g}, appear
to be quite successful in the lower $pf$-shell ($A \leq$ 52).
But these modifications turn out to be insufficient
 for a consistent description of the cross-shell properties: e.g., 
 the 2$^+_1$ level of $^{56}$Ni with KB3G
 is predicted about 2 MeV higher than the observed value.

The FPD6 interaction\cite{fpd6} is of another type:  
an analytic two-body potential was assumed with parameters 
determined by a fit to the experimental data of $A$=41$\sim$49 nuclei.
It successfully describes heavier $pf$-shell nuclei,
 such as $^{56}$Ni and $^{64}$Ge \cite{qmcd,mcsm}.
There are, however, some defects, for instance, 
in the single-particle aspects of $^{57}$Ni.

We now turn to our new interaction.
An effective interaction for the $pf$-shell can be specified uniquely 
in terms of interaction parameters consisting of 
4 single-particle energies $\epsilon_a$
 and 195 two-body matrix elements $V_{JT}(ab;cd)$,
 where $a, b, \cdots$ denote single-particle orbits,
 and $JT$ stand for the spin-isospin quantum numbers.
Although $\epsilon_a$'s include kinetic energies as well,
they are treated as a part of the effective interaction as usual.  
We adjust the values of the interaction parameters
so as to fit experimental binding energies and energy levels.
We outline the fitting procedure here, while
details can be found in \cite{gf40a}.
For a set of $N$ experimental energy data $E_{\rm exp}^k$
 ($k=1$, $\cdots$, $N$),
 we calculate corresponding shell-model eigenvalues $\lambda_k$'s.
 We minimize the quantity
 $\chi^2 = \sum_{k=1}^N ( E_{\rm exp}^k - \lambda_k )^2$
 by varying the values of the interaction parameters. 
Since this minimization is a non-linear process with respect to
the interaction parameters, we solve it in an iterative way
with successive variations of those parameters followed by
diagonalizations of the Hamiltonian until convergence.

Experimental energies used for the fit are limited to those of ground 
and low-lying states.  Therefore, certain linear combinations (LC's) of 
interaction parameters are sensitive to those data and can be well determined, 
whereas the rest of the LC's are not.
We then adopt the so-called LC method\cite{lc}, where the well-determined
LC's are separated from the rest:
starting from an initial interaction,
 well-determined LC's are varied by the fit,
 while the other LC's are kept unchanged
 (fixed to the values given by the initial interaction).

In order to obtain shell-model energies, both the conventional and MCSM 
calculations are used.  Since we are dealing with global features of the
low-lying spectra for essentially all $pf$-shell nuclei, 
we use a simplified version of MCSM: 
we search a few (typically three) most important basis states 
(deformed Slater determinants) for each spin-parity, 
 and diagonalize the Hamiltonian matrix in the few-dimensional
basis approximation (FDA).
The energy eigenvalues are improved by an empirical correction formula 
with parameters fixed by cases where 
more accurate results are available.  This is called hereafter
few-dimensional approximation with empirical corrections (FDA*), which
actually yields a reasonable estimate of the energy eigenvalues 
with much shorter computer time.

In the selection of experimental data, 
 in order to eliminate intruder states
 from outside the present model space,
 we consider nuclei of $A \geq 47$ and $Z \leq 32$.
As a result 699 data of binding and excitation energies
 (490 yrast, 198 yrare and 11 higher states) were taken
 from 87 nuclei : $^{47-51}$Ca, $^{47-52}$Sc, $^{47-52}$Ti, $^{47-53,55}$V,
 $^{48-56}$Cr, $^{50-58}$Mn, $^{52-60}$Fe, $^{54-61}$Co, $^{56-66}$Ni,
 $^{58-63}$Cu, $^{60-64}$Zn, $^{62,64,65}$Ga and $^{64,65}$Ge.
We assume an empirical mass dependence $A^{-0.3}$
 of the two-body matrix elements similarly to the USD interaction\cite{usd}.
We start from the realistic G-matrix interaction
 with core-polarization corrections based on 
the Bonn-C potential \cite{g-mat}, which is simply denoted G hereafter.
70 well-determined LC's are varied in the fitting procedure, 
and a new interaction, GXPF1, was obtained 
 with a rms error 168 keV within FDA*.

\begin{figure}
\includegraphics[scale=0.45]{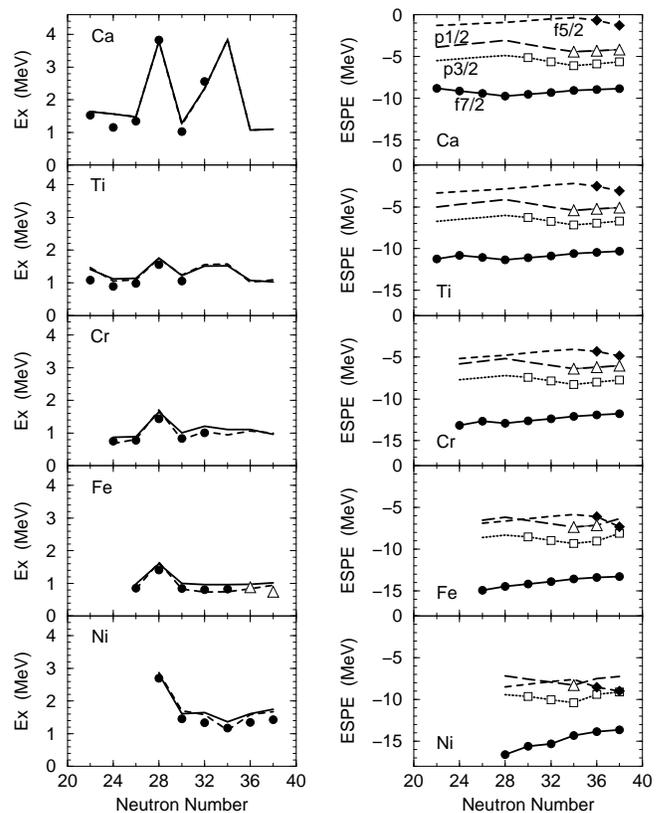}
\caption{(Left) First 2$^+$ levels as a function of
 the neutron number $N$. 
Experimental data are shown by filled circles
 \protect\cite{toi} and open triangles \protect\cite{hanna}.
Solid lines show results of conventional calculations: 
the maximum number of nucleons excited
 from $f_{7/2}$ to $p_{3/2}$, $p_{1/2}$ or $f_{5/2}$
 is 5 for $^{56}$Fe and $^{58,60,62}$Ni,
 6 for $^{52,54}$Fe and $^{56}$Ni, and 7 for $^{58,60}$Fe, whereas
the others are exact.
Dashed lines imply FDA* results.
(Right) Effective single-particle energies
 for neutron orbits. Symbols indicate that
 the corresponding orbit is occupied by at least one nucleon
 in the lowest filling configuration.}
\label{fig:2plus}
\end{figure}

The first $2^+$ energy level of even-even nuclei are
 a good systematic measure of the structure.
The left panel of Fig. \ref{fig:2plus} shows them for the
 Ca, Ti, Cr, Fe, and Ni isotopes.
The lightest nucleus in each isotope chain
 corresponds to $N$=$Z$ because of the mirror symmetry.
The energies are computed by FDA* and by  
exact or nearly exact conventional shell-model calculations 
by the code MSHELL\cite{mshell}.
The former gives a reasonable approximation to the latter, 
while the differences are up to 0.2 MeV.
The overall description of the $2^+$ levels 
 is quite successful throughout these isotope chains.
In all cases, the energy jump corresponding to $N=28$ shell closure
 is nicely reproduced.

\begin{figure*}
\includegraphics[scale=0.69]{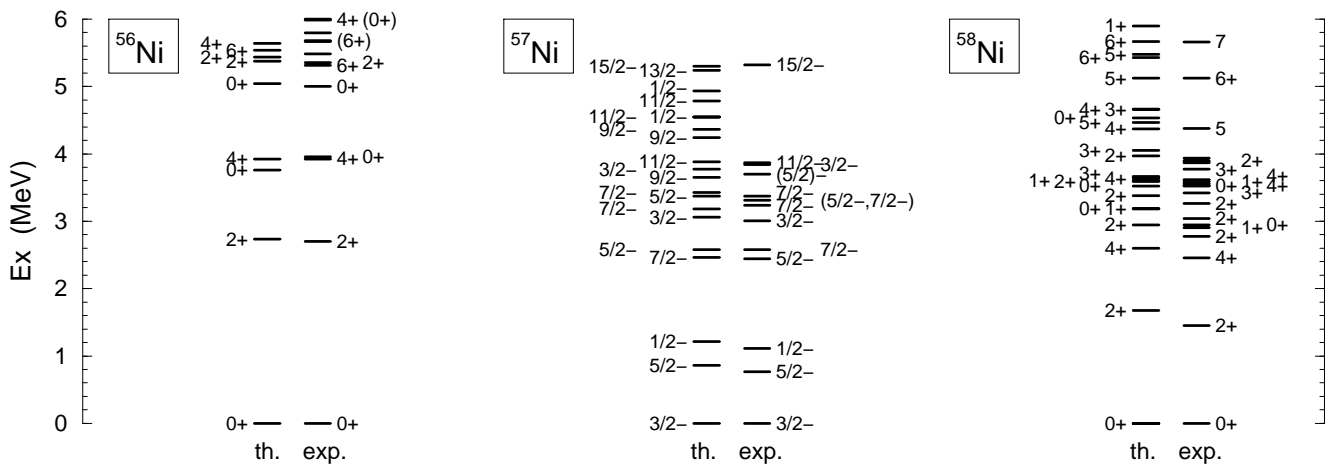}
\caption{Energy levels of $^{56,57,58}$Ni.
 Experimental data are taken from ref.\protect\cite{toi}.
Above 4 MeV, experimental levels are shown only for yrast states
 for $^{57,58}$Ni.}
\label{fig:ni}
\end{figure*}

A basic aspect of the effective interaction is provided by 
 the effective single-particle energies (ESPE)\cite{kb3,mcsm-n20}.  
The ESPE depends on the monopole part of the Hamiltonian,
 and reflects angular-momentum-averaged effects of the two-body interaction
for a given many-body system.  In the right panel of Fig. \ref{fig:2plus},
 ESPE's of the neutron orbits are shown.
A new $N$=34 magic structure has been predicted in \cite{magic}.
In fact a large energy gap ($\sim$ 4.1 MeV) 
 between $p_{1/2}$ and $f_{5/2}$ can be seen in the ESPE for Ca and Ti 
isotopes.  

As predicted also in \cite{magic}, due to the 
large attractive proton-neutron matrix elements, the ESPE of the 
 $\nu f_{5/2}$ orbit comes down as $\pi f_{7/2}$ orbit is occupied,
 reducing this subshell gap.  One can indeed see this change 
in Fig. \ref{fig:2plus}:  the $N$=34 subshell becomes weaker in Cr, and 
 disappears in Fe and Ni.  This means that the $N$=34 magic number arises
only in neutron-rich Ca and Ti isotopes, and is gone in stable nuclei.
  On the other hand, the ESPE's of Ni isotopes in Fig. \ref{fig:2plus} 
  indicate that the $f_{5/2}$, $p_{3/2, 1/2}$ orbits are degenerate to a 
  good extent, forming a subshell which resembles a degenerate pseudo $sd$-
  shell.  This fact may be relevant in constructing some algebraic models, 
  for instance, a pseudo SU(4) of IBM-4 \cite{pseudo_su4}.

In the Ca isotopes, a prominent peak in the calculated 2$^+$ excitation
energy can be seen
 at $N$=34, which is as high as that of the doubly magic nucleus $^{48}$Ca.
This is exactly due to the $N$=34 subshell closure discussed above.
Note that this calculation is a result of the diagonalization of the 
Hamiltonian, whereas the ESPE reflects only its monopole part. 
The subshell gap at $N$=32 is much smaller ($\sim$1.6 MeV) 
between $p_{3/2}$ and $p_{1/2}$, and shows less pronounced effect
in a consistent manner with experiment.
The gap at $N=34$ is not large with the FPD6 interaction
as can be inferred from the 2$^{+}$ state systematics shown in Fig. 4
of \cite{prisc}. Thus the experimental energy of the 2$^{+}$ state
in $^{54}$Ca will be an important test of the $pf$-shell Hamiltonians.

In the Ni isotopes, both experimental and theoretical $2^+$
 excitation energies drop at $N$=34,
 where ESPE of $\nu p_{1/2}$ and $\nu f_{5/2}$ are almost degenerate
 and therefore the collectivity is enhanced.
It is remarkable that the drastic change of the structure among those
nuclei can be described by a single effective interaction.
Toward the end of the $pf$-shell ($N$=40), effects of $g_{9/2}$ seem to
appear.

The nucleus $^{56}$Ni is a challenge for a unified description of nuclei in the
cross-shell region of the $pf$-shell.
The stability of the $(f_{7/2})^{16}$ 
core plays important roles, while 
most of existing effective interactions fail in reproducing certain 
properties related to the core softness.
In Fig. \ref{fig:ni} energy levels of $^{56,57,58}$Ni are shown.
Theoretical levels are obtained by MCSM calculations, where
 typically 13 $J$-compressed basis states\cite{mcsm} are taken for each state.
These MCSM calculations are more accurate than the FDA*.
The agreement between the theory and the experiment is satisfactory.

In the calculated spectrum of $^{56}$Ni,
 the deformed band discussed in \cite{ni56deform,rudolph}
 appears as $0^+_3$, $2^+_2$, $4^+_4$, $6^+_3$, $\cdots$.
In the yrast band, we obtain
 $B$(E2; $0^+_1$ $\rightarrow$ $2^+_1$)=5.5$\times$10$^2$ e$^2$fm$^4$,
 while in the deformed band,
 $B$(E2; $0^+_3$ $\rightarrow$ $2^+_2$)=1.8$\times$10$^3$ e$^2$fm$^4$.
Here, effective charges, $e_p$=1.23, $e_n$=0.54, are taken \cite{mcsm}. 
The former $B$(E2) value is in agreement with experiment \cite{kraus}.
The probability of the $(f_{7/2})^{16}$ configuration in the
 ground state is 69\%.  This value is larger than that
 obtained by FPD6 (49\%)\cite{mcsm}, while is smaller than 
the corresponding quantity for $^{48}$Ca (94\% for GXPF1).

In order to study the core-softness,
 we made a rather ``soft'' interaction (GXPF2).
For 20 matrix elements of the type
 $V_{JT}(a a; b c)$ where $a=f_{7/2}$ and $b, c \neq a$,
we keep their G interaction values.  We then carried out a fit for  
 60 best determined LC's, and came up with 
 a rms error of 188 keV for 623 data within FDA*.
This means that the fit is only slightly worse (by $\sim$20 keV) than GXPF1.
The probability of $(f_{7/2})^{16}$ configuration in the ground state 
of $^{56}$Ni is 49\% suggesting its softness.
Thus the energy data included in the fit can be reproduced about equally 
well within certain allowance of the softness.
The core property, however, is important for certain 
 observables.  For example, the total Gamow-Teller strength from the
 ground state is calculated as $B(GT_+)$=11.3 (9.5) for
 GXPF1 (GXPF2).  Since this is 13.7 for the closed 
shell, one sees the degree of the quenching which can be crucial in the
electron capture in the supernovae explosion.  From the KB3, one obtains
10.1 \cite{langGT}. 

In the lowest three states $3/2^-$, $5/2^-$, and $1/2^-$ of $^{57}$Ni,
the $(f_{7/2})^{16}$ core is broken similarly to $^{56}$Ni ground state, and
these are ``single-particle'' states built on top of the correlated 
$^{56}$Ni ground state. 
Their relative positions are determined mainly by the ESPE's.
In $pf$-shell nuclei, in general, the cross-shell excitations occur rather 
commonly in low-lying states \cite{semi}.
In $^{57}$Ni, states above the three ``single-particle'' states
contain further cross-shell excitations, 
which provide a good testing ground 
for cross-shell two-body matrix elements.  
A similar situation has been seen
in $^{56}$Ni particularly for its 4p-4h
deformed band, while such effects are  
less evident in $^{58}$Ni, with an exception of the $0^+_3$ state 
which contains a sizable amount of proton 2p-2h 
excitations \cite{nakada}. In these examples np-nh refers to the
particle-hole excitations 
across the gap on top of the correlated 
ground state \cite{semi}.
A comprehensive picture for such a wide variety of 
states requires the
full-space calculations with an appropriate effective interaction.

\begin{figure}
\includegraphics[scale=0.62]{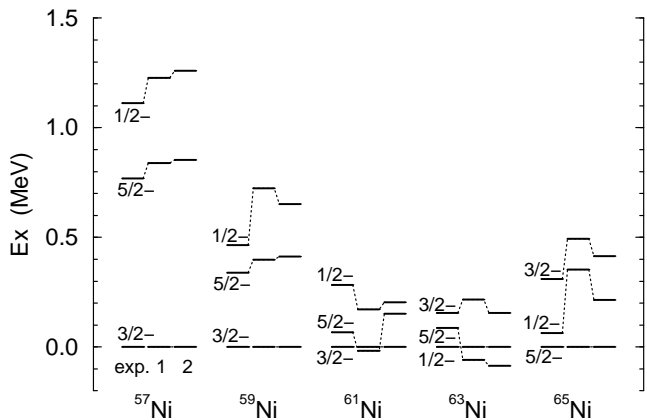}
\caption{
Energy levels of odd-$A$ Ni isotopes.
Experimental data are taken from ref.\protect\cite{toi}.
Theoretical results are calculated by using GXPF1/2
 interactions, which are indicated as 1 and 2,
 respectively. 
Shell model calculations
 were carried out by the code MSHELL\protect\cite{mshell}.
The results for $^{57-63}$Ni are obtained with truncating the number of
nucleons excited from $f_{7/2}$ orbit up to six, whereas the result
for $^{65}$Ni is exact.}
\label{fig:odd}
\end{figure}

The description of odd-$A$ nuclei is also an important test of the
 interaction.
We again consider Ni isotopes as examples, for which 
 energy levels of yrast
 $1/2^-$, $3/2^-$, and $5/2^-$
 states are shown in Fig.\ref{fig:odd}.
Experimentally, as the number of neutrons increases,
 both $1/2^-$ and $5/2^-$ states come down
 relative to $3/2^-$ level,
 forming nearly degenerate states around $^{63}$Ni.
This characteristic feature is nicely reproduced by both GXPF1/2 interactions.
The downward slope of $5/2^-$ level, however, 
 seems to be somewhat too steep, especially in GXPF1, which leads to
 some deviations within about 0.2 MeV, while the basic degeneracy
 around $^{63}$Ni is clearly maintained.
Since the results of the FDA* used in the fit are
 much closer to experimental spectra,
 a part of this deviation is due to the
 uncertainty in the FDA*.

These yrast levels contain similar amount of cross-shell excitations
 within each isotope.
In fact, the occupation number of $f_{7/2}$ orbit varies
 from 15.0 (14.6) for $^{57}$Ni to 15.6 (15.5) for $^{65}$Ni by GXPF1 (GXPF2),
 while the difference among the lowest 3 states within each
 isotope is less than 0.3.
Therefore the core excitation appears to play a minor role for describing
 relative energies of these lowest 3 levels.
However, it should be emphasized that
 the GXPF1/2 interactions can describe these
 lowest levels as well as non-yrast levels
 for which the cross-shell excitation is essential.

In usual shell model interactions,
 we often take the single-particle energies $\epsilon_a$
 from experimental
 energy spectra of the one-particle/hole system
 on top of the assumed inert core.
In the case of $pf$-shell, many existing effective interactions
 such as KB3 borrow them from $^{41}$Ca.
However, our main purpose is
 not to describe light $pf$-shell nuclei from the beginning of the shell, 
 but to treat cross-shell excitations over $N$ or $Z=28$ shell gap.
Therefore, in the present approach, we have assumed single-particle energies
 as parameters and determined them by the fit.

In Table \ref{tbl:spe} the single particle energies relative to
 $p_{3/2}$ orbit are shown for various interactions.
It is remarkable that the single-particle energy spacing
 between $f_{7/2}$ and $p_{3/2}$ orbits in GXPF1/2 interactions
 is enhanced by about 0.9 MeV
 in comparison to the energy spectrum of $^{41}$Ca (or KB3).
This difference in single-particle energies results from the exclusion
 of energy data of $A<47$ nuclei from the fit. 
The wave functions of nuclei
 near $^{40}$Ca contain relatively large amounts of intruder state
 admixtures, and the $sd$-shell should be included in their description. 
A fit in which the single-particle energies are fixed to their values in
 $^{41}$Ca is possible, but the total rms error in the energies becomes
 larger. 
The ESPE of GXPF1 interaction for Ca isotopes becomes very close
 to that of KB3, which is known to be quite successful for light
 $pf$-shell nuclei, already around $A\sim 45$ (where the difference is
 less than 0.2 MeV). 
Thus the description of $A<47$ nuclei by GXPF1/2
 interactions is still reasonable and similar to that of KB3, except for
 the those nuclei at the very beginning of the shell. 

\begin{table}
\caption{Single-particle energies relative to 
$p_{3/2}$ of various interactions.}
\label{tbl:spe}
\begin{ruledtabular}
\begin{tabular}{lrrrr}
orbit & KB3 & FPD6 & GXPF1 & GXPF2 \\ \hline
$f_{7/2}$ & $-$2  &  $-$1.8924 & $-$2.9447 & $-$2.8504 \\
$p_{1/2}$ &  2    &   2.0169   &   1.5423  &   1.6054 \\
$f_{5/2}$ & 4.5   &   4.5986   &   4.2964  &   4.2584
\end{tabular}
\end{ruledtabular}
\end{table}

\begin{figure}
\includegraphics[scale=0.81]{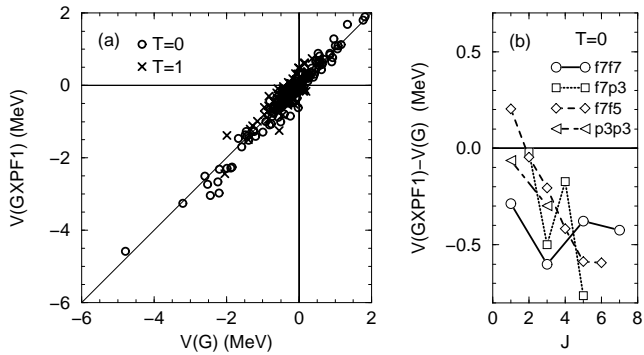}
\caption{(a) Correlation of $V_{JT}(ab;cd)$ between G and GXPF1.
$T$=$0$ and $T$=$1$ matrix elements are shown by open circles and crosses,
 respectively.
(b) Difference of diagonal $T$=$0$ matrix elements
 between G and GXPF1 as a function of the angular momentum, $J$, coupled by 
 two nucleons.  }
\label{fig:gxpf1}
\end{figure}

In Fig. \ref{fig:gxpf1}(a), a comparison between GXPF1 and G 
is shown
for the 195 two-body matrix elements.  One finds a strong correlation. 
 On average, the $T$=0 ($T$=1) matrix elements are modified to be more 
attractive (repulsive). 
The majority of most attractive matrix elements are $T$=0 ones, and, 
in particular, the two most attractive ones belong to $T$=0 f7f5 in
both GXPF1 and G, where the notation f7f5 refers to the set of
matrix elements $V_{J,T}(ab;ab)$ with $a=f_{7/2}$, $b=f_{5/2}$.

In Fig. \ref{fig:gxpf1}(b) the difference
 between GXPF1 and G is shown for several $T$=0 diagonal matrix elements.
The fit makes them more attractive. 
Although the f7f7 matrix elements are dominated by the monopole shift,
i.e., a $J$-independent correction, 
 the corrections to cross-shell matrix elements f7p3 and f7f5,
 become larger for higher $J$'s.
Such $J$-dependences beyond the monopole shifts are of interest. 
The same feature is also found in GXPF2.
Apart from these corrections, notable differences between G and GXPF1
 are found in the $T=1$ monopole-pairing cross-shell matrix elements.

In summary, a new unified $pf$-shell effective interaction, GXPF1, 
has been obtained. The GXPF1 interaction
has monopole properties enforced by the energy data, 
and it properly handles
cross-shell excitations, leading to a successful description of structure
of Ni isotopes. 
Corrections beyond the monopole shifts are important.
The systematic behavior of the energies for the lowest $2^+$ levels of
even-even  nuclei are in good agreement with experiment, suggesting that
collective properties are well described. 
We investigated the possible range of the softness of $N$=$Z$=28
core, coming up with another interaction, GXPF2.
The applications to unexplored regimes of large neutron numbers or
high excitation energy is of great interest.
For example,   
the GXPF1 interaction demonstrates the appearance of a new magic number 
$N$=34 \cite{magic} in neutron-rich nuclei.  
Future experiments
will test the predictions
and provide guidance for further improvements in the Hamiltonian.

The authors thank M. Hjorth-Jensen for providing us the interaction matrix elements.
A part of numerical calculations were carried out by a
 parallel computer, Alphleet at RIKEN.  
This work was supported in part by 
a Grant-in-Aid for Scientific Research (A)(2)(10304019), and by a Grant-in-Aid
for Specially Promoted Research (13002001), and by the US National
Science Foundation Grant PHY-0070911.

\end{document}